\documentclass[12pt,aps,showpacs]{revtex4}
\usepackage{amsmath,amssymb,amsfonts,epsfig}
\newcommand{\beeq}{\begin{equation}}
\newcommand{\eneq}{\end{equation}}
\newcommand{\be}{\begin{eqnarray}}
\newcommand{\ee}{\end{eqnarray}}
\newcommand{\bpic}{\begin{picture}}
\newcommand{\epic}{\end{picture}}
\newcommand{\bs}{\begin{scriptsize}}
\newcommand{\es}{\end{scriptsize}}

\def\la{\raise.16ex\hbox{$\langle$} \, }
\def\ra{\, \raise.16ex\hbox{$\rangle$} }

\def\Box{\kern1pt\vbox{\hrule height 1.2pt\hbox{\vrule width 1.2pt\hskip 3pt
   \vbox{\vskip 6pt}\hskip 3pt\vrule width 0.6pt}\hrule height 0.6pt}\kern1pt}
\def\gtwid{\mathrel{\raise.3ex\hbox{$>$\kern-.75em\lower1ex\hbox{$\sim$}}}}
\def\ltwid{\mathrel{\raise.3ex\hbox{$<$\kern-.75em\lower1ex\hbox{$\sim$}}}}

\begin{document}


\title{Non-perturbative results for a massive scalar field\\during de Sitter inflation }

\author{G. Karakaya}\email{gulaykarakaya@itu.edu.tr}
\affiliation{Department of Physics, Istanbul Technical
University, Maslak, Istanbul 34469, Turkey}
\affiliation{Department of Industrial Engineering, Kadir Has University, 34083, Istanbul, Türkiye.}

\begin{abstract}
We consider an infrared truncated massive minimally coupled scalar field with an asymmetric self-interaction $\frac{m^2}{2}\varphi^2\!+\!\frac{\lambda\varphi^4}{4!}\!+\!\frac{\beta\varphi^3}{3!}(\lambda\!>\!0)$ during a cosmological constant driven de Sitter inflation with a constant expansion rate, $H$. The Fokker-Planck equation for the evolution of the massive scalar field is obtained and solved for this asymmetric potential. Firstly we compute the vacuum expectation values of $\varphi$ and $\varphi^2$ using the normalized late-time probability distribution $\rho(\varphi,t)$. Secondly, we evaluate the two-point correlation function and the vacuum expectation value of the massive scalar field at tree and one-loop order following Starobinsky's approach and applying the techniques of perturbative quantum field theory. Lastly, we compare the results obtained via these two different methods. Although these results give consistent qualitative behavior at tree and one-loop order, they differ numerically.
\end{abstract}

 
\maketitle \vskip 0.2in \vspace{.4cm}

\section{Introduction}
\label{sec:intro}

Two-point correlation function of an infrared (IR) truncated {\it massive} minimally coupled scalar field with 
$\frac{m^2}{2}\varphi^2\!+\!\frac{\lambda\varphi^4}{4!}$  on a locally de Sitter background of an inflating spacetime was computed at tree, one-and two-loop order in Ref.~\cite{vacuum}. In this paper, we analytically evaluate the quantum corrected vacuum expectation value and two-point correlation function of the infrared(IR) truncated massive scalar field with $\frac{m^2}{2}\varphi^2\!+\!\frac{\lambda\varphi^4}{4!}\!+\!\frac{\beta\varphi^3}{3!}$  asymmetric potential using two different methods. 

However, perhaps the most effective formalism to resum these secular logarithms for scalar field theories without derivative interactions is the stochastic approach. It is employed for calculating expectation values related to the long-wavelength (infrared component) of the scalar field. This technique yields the secular logarithms at perturbative orders.
                    
Fluctuations of interacting scalar fields in an inflating spacetime have been the focus of cosmologists \cite{sfl}. Recently, there has been a revival of interest on IR dynamics of scalar potential models with various approaches that include extending the stochastic formalism \cite{TT}, applying complementary series analysis \cite{AMPP}, computing effective actions \cite{MR,R1,DB1,CW1}, using Schwinger-Keldish formalism \cite{CWX}, implementing Fokker-Planck equation and $\delta N$ formalism \cite{AFNVW}, employing $1/N$ expansion \cite{GS1}, adopting reduced density matrix method \cite{DB2}, applying renormalization group analysis \cite{GS2} and computing effective potentials \cite{JSS}. Influence of fermions on scalar field fluctuations has been studied in Refs~\cite{O2,DB3}. In this paper we use two different methods: Starobinsky's approach (Perturbative quantum field theory)  and Fokker-Planck equation. We consider an infrared truncated massive minimally coupled scalar field with an asymmetric self-interaction on a locally de Sitter background. The model is of interest because it exhibits, in the {\it massless} limit, peculiar {\it enhanced} quantum effects: the renormalized energy density and pressure of the scalar violate \cite{OW1,OW2} the weak energy condition on cosmological scales at two-loop order and a phase of superacceleration is induced. As the inflationary particle production amplifies the field strength and therefore forces the scalar up its potential, the scalar develops \cite{BOW} a positive self-mass squared which, in turn, reduces the particle production. Furthermore, the classical restoring force pushes the scalar back down to the configuration where the potential is minimum. Thus, the scalar cannot continue to roll up its position and comes to a halt eventually. The process, therefore, is self-limiting and the model is stable \cite{KO}. For many quantum field theory computations---involving ultraviolet modes---in cosmology higher order quantum corrections necessarily involve changes in the initial state. Neglecting to correctly change the initial state can result in effective field equations that diverge on the initial value surface. The model provides \cite{KOW1} an example of how perturbative initial state corrections can absorb initial value divergences. Moreover, the scalar makes \cite{KOW2} a time-dependent contribution to the amplitudes of curvature fluctuations at two-loop order. The amplitudes of the scalar field fluctuations grow \cite{O1} toward larger scales.

Two-point correlation function, $\langle\Omega|
\bar{\varphi}(t,\vec{x})\bar{\varphi}(t'\!,\vec{x}\,')|\Omega\rangle$, that we evaluate in this paper is an average measure of how the amplitude of the field at one event (spacetime coordinate) is correlated with the amplitude at another event. When the field strength grows, due to the inflationary particle production, so does the two-point correlation function. In fact, the growth is logarithmic in the massless limit, just as expected. As the mass increases, logarithmic growth must be suppressed.

In this work, we consider an infrared truncated massive minimally coupled scalar field during de Sitter inflation with an asymmetric self interaction potential
\begin{eqnarray}
V\left(\varphi\right)\!=\!\frac{m^2}{2}\varphi^2\!+\!\frac{\lambda\varphi^4}{4!}\!+\!\frac{\beta\varphi^3}{3!}\;. \nonumber
\end{eqnarray}

Vacuum decay in flat spacetime with the potential $V(\varphi)$ given above was studied long ago in \cite{Ven} for a 
massless and massive scalar, where mass generation and stability condition for the initial vacuum state were derived. In \cite{TW}, the same model was explored using the in-in formalism and a modified approach. The temporal evolution of the vacuum expectation value of the scalar field up to two-loop order, corresponding to cubic self-interaction, was explicitly demonstrated. This temporal behavior appears to differ from the pure logarithmic secular growth found in inflationary contexts, because of the lack of a natural length scale associated with Minkowski spacetime. Following this, the critical time at which perturbation theory might break down was estimated. For further discussion on the relevance of such a potential with a non-zero scalar mass in various inflationary models, we direct the reader to \cite{Ven}, inspired by WMAP data.

The hybrid potential for a massless minimally coupled scalar in the primordial inflationary scenario was first explored recently in \cite{Bhattacharya}. The rationale behind this choice is as follows. First, since the potential is bounded from below regardless of the sign or magnitude of $\beta$, it is expected to lead to a late-time equilibrium state, thus avoiding the problem of eternal rolling. we now assume that the system initially resides around $\varphi \sim 0$. Over time, the system will evolve towards the minima of the potential and eventually settle into these minima at late times. However, during this process, we anticipate strong non-perturbative radiative effects, arising from the secular contributions generated by the loops of the massless, minimally coupled scalar field. Clearly, these significant quantum effects must be accounted for to derive any meaningful conclusions about the final state of the system. Results from flat space-time alone may not provide accurate predictions due to these effects, which are uniquely associated with the inflationary scenario.

The outline of the paper is as follows. In Sec.~\ref{sec:model} we present the background geometry and the Lagrangian of the model. In Sec.~\ref{sec:fokker} we derive Fokker-Planck equation for the infrared truncated massive minimally coupled scalar field. Then, we calculate non-perturbative $\langle\bar{\varphi}\rangle$ and $\langle\bar{\varphi}^2\rangle$ for both of massive and massless scalar fields. We plot these expectation values versus cubic coupling parameter, $\beta$. In Sec.~\ref{sec:stochastic} we analyze the model following Starobinsky's approach and applying the techniques of perturbative quantum field theory. Then we compute the quantum corrected two-point correlation function and the vacuum expectation value of the IR truncated massive scalar in our model at tree and one-loop order.  We summarize our conclusions in Sec.~\ref{sec:conclusions}.
\section{The Model}
\label{sec:model}
We consider a massive, minimally coupled, self-interacting spectator scalar field during de Sitter inflation. The invariant line element 
\begin{eqnarray} 
ds^2\!\!=\! g_{\mu\nu} dx^{\mu  }
	dx^{\nu}\!\!=\!-dt^2\!\!+\!a^2(t) d\vec{x} \cdot d\vec{x}
	\; ,\label{desitter}\end{eqnarray} where the scale factor with a constant expansion rate $H$ is $a(t)\!=\!e^{H t}$. We work in a $4$-dimensional space-time. We adopt the convention in which a Greek index $\mu\!=\!0,1,2,3$, hence $x^\mu\!=\!(x^0\!,\vec x)$, $x^0\!\equiv\!t$, and $\partial_\mu\!=\!(\partial_0,\vec\nabla)$. 

We first evaluate the renormalized lagrange density of the model. The renormalized field is defined as $\varphi(x)\!\!\equiv\!\!\frac{1}{\sqrt{Z}}\phi(x)$. The bare lagrange density is given as 
\begin{eqnarray}
\mathcal{L}\!=\!-\frac{1}{2}\partial_{\mu}\phi\partial_{\nu}\phi g^{\mu\nu}
\sqrt{\!-\!g}\!-\!\frac{1}{2}m_{0}^2\phi^2\sqrt{\!-\!g}\!-\!\frac{\lambda_{0}}{4!}\phi^4\sqrt{\!-\!g}\!-\!\frac{\beta_{0}}{3!}\phi^3\sqrt{\!-\!g}\;,\label{lagrange}
\end{eqnarray}
where $m_0^2$ is the bare mass squared, $\lambda_0$ is the bare coupling constant, $\beta_{0}$ is the cubic coupling constant and $g$ is the determinant of the metric.
When we represent the bare parameters in terms of the renormalized parameters, they are expressed as
\begin{eqnarray}
 Z\!=\!1\!+\!\delta Z, \;Z m_{0}^2 
    \!=\!m^2\!+\!\delta m^2, \;Z^2\lambda_{0}\!=\!\lambda\!+\!\delta\lambda\; \text{and}\;Z^{\frac{3}{2}}\beta_{0}\!=\!\beta\!+\!\delta\beta\;.
\end{eqnarray}
The renormalized lagrangian density in our model becomes
\begin{eqnarray}
&&\hspace{1cm}\mathcal{L}\!=\!-\frac{1}{2}\partial_{\mu}\varphi\partial_{\nu}\varphi g^{\mu\nu}\sqrt{\!-\!g}\!-\!\frac{\lambda}{4!}\varphi^4\sqrt{\!-\!g}\!-\!\frac{\beta}{3!}\varphi^3\sqrt{\!-\!g}\\ \nonumber
&&\hspace{0.2cm}\!-\frac{\delta Z}{2}\partial_{\mu}\varphi\partial_{\nu}\varphi\sqrt{\!-\!g}\!-\!\frac{1}{2}\delta m^2\sqrt{\!-\!g}\!-\!\frac{\delta\lambda}{4!}\varphi^4\sqrt{\!-\!g}\!-\!\frac{\delta\beta}{3!}\varphi^3\sqrt{\!-\!g}\;,
\end{eqnarray}
where $\varphi(x)$ represents the renormalized  field and $m$ denotes the renormalized mass. The counterterms in the second line of the above equation will be neglected. They don't contribute for the IR field theory. 
\section{Quantum corrected correlators via Fokker-Planck equation}
\label{sec:fokker}
In this section, we calculate $\langle\bar{\varphi}^2\rangle$ via the Fokker-Planck equation for the infrared truncated massive scalar field, $\bar{\varphi}$. The Fokker-Planck equation is used as
\begin{eqnarray}
\frac{\partial\rho}{\partial t}\!=\!\frac{H^3}{8\pi^2}\frac{\partial^2\rho}{\partial\bar{\varphi}^2}\!+\!\frac{1}{3H}
\frac{\partial}{\partial\bar{\varphi}}\Big(\frac{\partial V}{\partial\bar{\varphi}}\rho\Big)\;,\label{Fokker-Planck}
\end{eqnarray}
where $\rho\!=\!\rho(\bar{\varphi},t)$ is the one-point probability density distribution function. The general solution of Eq. (\ref{Fokker-Planck}) is
\begin{eqnarray}
\rho(\bar{\varphi},t)\!=\!e^{\!-\!\nu(\bar{\varphi})}\sum_{n\!=\!0}^{\infty}a_n\Phi_{n}(\bar{\varphi})
e^{\!-\!\Lambda_n(t\!-\!t_{0})}\;,\label{kemal1}
\end{eqnarray}
where $t_0$ is some initial time, $\nu(\bar{\varphi})\!=\!4\pi^2V(\bar{\varphi})/3H^4$, $a_n$'s are coefficients independent of time and $\Phi_n(\bar{\varphi})$, $\Lambda_n$ are respectively the eigenfunctions and eigenvalues corresponding to the Schr{\"o}ndinger like equation 
\begin{eqnarray}
\!-\frac{1}{2}\frac{d^2\Phi_n(\bar{\varphi})}{d\bar{\varphi}^2}\!+\!
\frac{1}{2}\Big[(\nu\,'(\bar{\varphi}))^2\!-\!\nu\,''(\bar{\varphi})\Big]\Phi_n(\bar{\varphi})\!=\!
\frac{4\pi^2\Lambda_n}{H^3}\Phi_n(\bar{\varphi})\;,\label{kemal}
\end{eqnarray}
The eigenfunctions $\Phi_n(\bar{\varphi})$'s satisfy the orthogonality condition
\begin{eqnarray}
\int d\bar{\varphi} \Phi_n(\bar{\varphi})\Phi_m(\bar{\varphi})\!=\!\delta_{nm}\;,
\end{eqnarray}
 which one can find out the coefficients $a_n$. Note also that Eq. $(\ref{kemal})$ can be rewritten in the form
\begin{eqnarray}
\frac{1}{2}\Bigg(\!-\!\frac{\partial}{\partial\bar{\varphi}}\!+\!\nu\,'(\bar{\varphi})\bigg)
\Bigg(\frac{\partial}{\partial\bar{\varphi}}\!+\!\nu\,'(\bar{\varphi})\bigg)\Phi_n(\bar{\varphi})\!=\!
\frac{4\pi^2\Lambda_n}{H^3}\Phi_n(\bar{\varphi})\;.
\end{eqnarray}
Here $\bar{\varphi}$ is real, so we have $(\partial_{\bar{\varphi}})^{\dagger}\!=\!-\partial_{\bar{\varphi}}$. Therefore $\Lambda_n$'s are eigenvalues of a positive operator of the form $A^{\dagger}A$, so $\Lambda_n\geq0$ with $\Lambda_{0}\!=\!0$ represent the ground state. The corresponding wave function is 
\begin{eqnarray}
\Phi_{0}(\bar{\varphi})\!=\!N^{\frac{\!-\!1}{2}}e^{\!-\frac{4\pi^2V(\bar{\varphi})}{3H^4}}\;,
\label{kemal3}
\end{eqnarray}
where $N$ is the normalisation, exists only if $V(\bar{\varphi})$ is bounded from below, in which case late time equilibrium state is possible. The equilibrium probability distribution is given Eqns. (\ref{kemal1}), (\ref{kemal3}) by
\begin{eqnarray}
\rho_{eq}(\bar{\varphi})\!=\!N^{\!-\!1}e^{\frac{\!-\!8\pi^2V(\bar{\varphi})}{3H^4}}\;.
\end{eqnarray}
The equilibrium probability distribution function allows us to compute the expectation value of any operator $\Xi(\bar{\varphi})$, at late times as 
\begin{eqnarray}
\langle\Xi(\bar{\varphi})\rangle:=\int_{-\infty}^{\infty}d\bar{\varphi}\Xi(\bar{\varphi})
\rho_{eq}(\bar{\varphi})\;.
\end{eqnarray}
From now  we can write $\bar{\varphi}$ simply as $\varphi$ in this section. The coincident correlator $\langle\varphi^2\rangle$ is 
 \begin{eqnarray}
 \langle\varphi^2\rangle\!=\!N^{\!-\!1}\int_{-\infty}^{\infty}d\varphi \varphi^2 e^{\frac{\!-\!8\pi^2}{3H^4}(\frac{1}{2}m^2\varphi^2\!+\!\frac{\lambda}{4!}\varphi^4\!+\!\frac{\beta}{3!}\varphi^3)}\;,\label{corr}
 \end{eqnarray}
 where
 \begin{eqnarray}
 N\!=\!\int_{-\infty}^{\infty} d\varphi e^{\frac{\!-\!8\pi^2}{3H^4}(\frac{1}{2}m^2\varphi^2\!+\!\frac{\lambda}{4!}\varphi^4\!+\!
 	\frac{\beta}{3!}\varphi^3)}\;.
 \end{eqnarray}

Using the power series expansion of exponential for the cubic potential, the above integral is rewritten as
\begin{eqnarray}
\langle\varphi^2\rangle\!=\!N^{\!-\!1}\int_{-\infty}^{\infty}d\varphi \varphi^2\sum_{n\!=\!0}^{\infty}\frac{1}{n!}\Big(\!-\!\frac{4\pi^2\beta\varphi^3}{9H^4}\Big)^n
 e^{\frac{\!-\!8\pi^2}{3H^4}(\frac{1}{2}m^2\varphi^2\!+\!\frac{\lambda}{4!}\varphi^4)}\;,
 \end{eqnarray}
 where
 \begin{eqnarray}
N\!=\!\int_{-\infty}^{\infty}d\varphi \sum_{n\!=\!0}^{\infty}\frac{1}{n!}\Big(\!-\!\frac{4\pi^2\beta\varphi^3}{9H^4}\Big)^n
 e^{\frac{\!-\!8\pi^2}{3H^4}(\frac{1}{2}m^2\varphi^2\!+\!\frac{\lambda}{4!}\varphi^4)}\;.
 \end{eqnarray}
 We find
 \begin{eqnarray}
 &&\langle\varphi^2\rangle\!=\!N^{\!-\!1}\sum_{n\!=\!0}^{\infty}
 \frac{2^{4n-1}\pi^{\frac{2n\!-\!3}{2}}\beta^{2n}}{(2n)!3^{\frac{2n\!-\!3}{2}}H^{2n-3}
 	\lambda^{\frac{6n\!+\!7}{4}}}\Bigg\{\lambda\Gamma\Big[\frac{3(2n\!+\!1)}{4}\Big]
 {}_{1}\mathcal{F}_{1}\left(\frac{3(2n\!+\!1)}{4};\frac{1}{2};
 \frac{4m^4\pi^2}{H^4\lambda}\right)\!-\!4\pi m^2\frac{\sqrt{\lambda}}{H^2}
 \nonumber\\
 &&\hspace{3cm}\times\Gamma\Big[
 \frac{5\!+\!6n}{4}\Big]{}_{1}\mathcal{F}_{1}\left(\frac{5\!+\!6n}{4};\frac{3}{2};
 \frac{4m^4\pi^2}{H^4\lambda}\right)
 \Bigg\}\;,\label{gulay1}
 \end{eqnarray}
 where
 \begin{eqnarray}
 &&N\!=\!\sum_{n\!=\!0}^{\infty}
 \frac{2^{4n\!-\!1}\beta^{2n}3^{\frac{(1\!-\!2n)}{2}}}{(2n)!
 	\pi^{\frac{1\!-\!2n}{2}}H^{2n\!-\!1}\lambda^{\frac{5\!+\!6n}{4}}}\Bigg\{\lambda\Gamma
 \Big[\frac{1\!+\!6n}{4}\Big]{}_{1}\mathcal{F}_{1}\left(\frac{1\!+\!6n}{4};\frac{1}{2};
 \frac{4m^4\pi^2}{H^4\lambda}\right)\!-\!4\pi m^2\frac{\sqrt{\lambda}}{H^2}
 \nonumber\\
 &&\hspace{3cm}\times\Gamma[\frac{3(2n\!+\!1)}{4}]{}_{1}\mathcal{F}_{1}\left(\frac{3(2n\!+\!1)}{4};\frac{3}{2};\frac{4m^4\pi^2}{H^4\lambda}\right)
 \Bigg\}\;.
 \end{eqnarray}
 \begin{figure}
 	\centering
  \includegraphics[width=11.5cm,height=6.5cm]{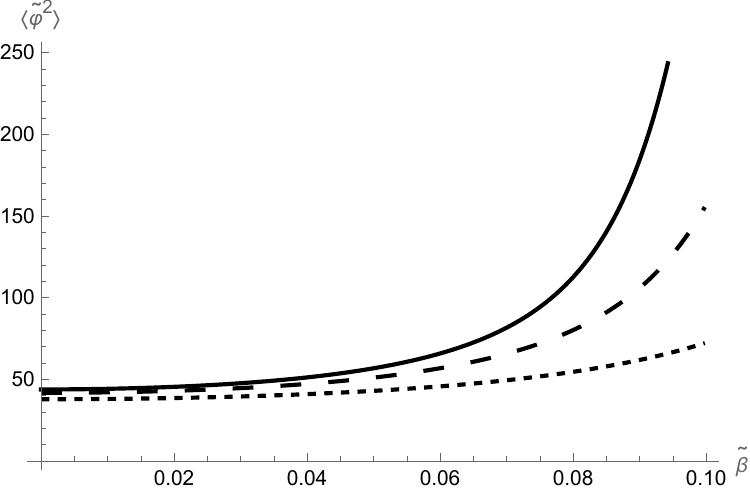}
 	\vspace*{2.256mm}
 	\caption{Plots of $\langle\tilde{\varphi}^2\rangle$ for the massive scalar, defined in Eq.~(\ref{gulay1}) versus $\tilde{\beta}$ for  different values of $\lambda$. The dashed, large-dashed and solid lines are for $\lambda\!=\!0.015, 0.01$ and $0.008$, respectively.}
 	\label{kemal6}
 \end{figure}
 In Fig$.\!\quad\!1$, as the coupling constant $\lambda$ increases, the coincident correlator of the massive scalar, 
$\langle \tilde{\varphi}^{2} \rangle$ decreases.  
Conversely, as the cubic coupling constant $\tilde{\beta}$ increases, the coincident correlator increases.

 In the massless limit, the equal space-time correlator \cite{Joshi} is
 \begin{eqnarray}
&&\hspace{1cm}\lim_{m\rightarrow0}\langle\varphi^2\rangle\!=\!\Bigg(6\lambda^{\frac{3}{2}}\Gamma\Big[\frac{3}{4}
 \Big]{}_{2}\mathcal{F}_{2}\Big[\Big\{\frac{7}{12},\frac{11}{12}\Big\},
 \Big\{\frac{1}{2},\frac{3}{4}\Big\},\frac{3\pi^2\beta^4}{H^4\lambda^3}\Big]\!+\!
 \frac{5\pi\beta^2}{H^2}\Gamma\Big[\frac{1}{4}\Big]
 \nonumber\\
 &&\times
 {}_{2}\mathcal{F}_{2}\Big[\Big\{\frac{13}{12},\frac{17}{12}\Big\},
 \Big\{\frac{5}{4},\frac{3}{2}\Big\},\frac{3\pi^2\beta^4}{H^4\lambda^3}\Big]\bigg)/\Bigg(
 2\pi\frac{\lambda^2}{H^2}\Gamma\Big[\frac{1}{4}\Big]{}_{2}\mathcal{F}_{2}\Big[\Big\{
 \frac{1}{12},\frac{5}{12}\Big\},
 \Big\{\frac{1}{4},\frac{1}{2}\Big\},\frac{3\pi^2\beta^4}{H^4\lambda^3}\Big]
 \nonumber\\
 &&\hspace{2cm}\!+\frac{4\pi^2\beta^2}{H^2}\sqrt{\frac{\lambda}{H^4}}\Gamma\Big[\frac{3}{4}\Big]
 {}_{2}\mathcal{F}_{2}\Big[\Big\{\frac{7}{12},\frac{11}{12}\Big\},
 \Big\{\frac{3}{4},\frac{3}{2}\Big\},\frac{3\pi^2\beta^4}{H^4\lambda^3}\Big]
 \bigg)\;.\label{gülay}
 \end{eqnarray}
As $\lambda$ increases, the equal space-time correlator decreases, whereas an increase in $\beta$ causes the correlator to increase.

 \begin{figure}
 	\centering
 	\includegraphics[width=11.5cm,height=6.5cm]{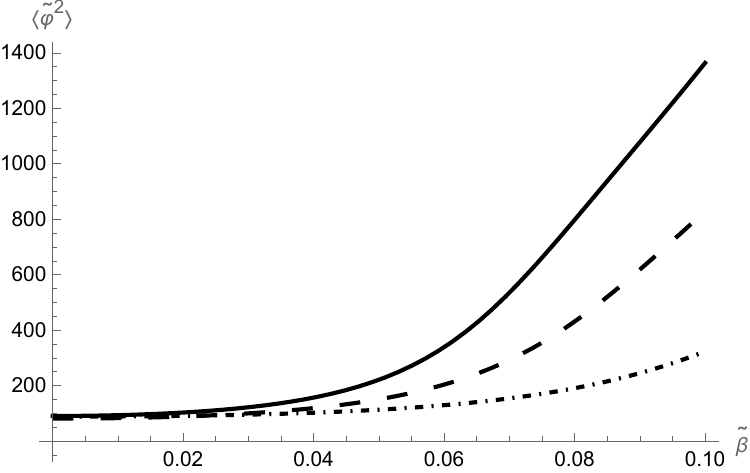}
 	\vspace*{2.256mm}
 	\caption{Plots of $\langle\tilde{\varphi}^2\rangle$ for the massless scalar, defined in Eq.~(\ref{gülay}) versus $\tilde{\beta}$ for  different values of $\lambda$. The solid, large-dashed and dashed lines are for $\lambda\!=\!0.008, 0.01$ and $0.015$, respectively.}
 	\label{gülay6}
 \end{figure}
There are the two special cases: small and large cubic coupling constants. Firstly,  we obtain the $\langle\varphi^2\rangle$ for $\tilde{\beta}\rightarrow0 $ as,
\begin{eqnarray}
\langle\varphi^2\rangle\!=\!H^2\Big(
\frac{0.3228}{\sqrt{\lambda}}\!+\!1.8146
\frac{\tilde{\beta^2}}{\lambda^2}\!+\!\mathcal{O}\Big(
\frac{\tilde{\beta^4}}{\lambda^{7/2}}\Big)\Big)\;,
\end{eqnarray}
where $\tilde{\beta}\!=\!\frac{\beta}{H}$
and for large $\tilde{\beta}^4/\lambda^3\gg1 $, Eq.~$\left(\ref{gülay}\right)$ is obtained as
\begin{eqnarray}
\langle\varphi^2\rangle\!=\!H^{2}\frac{\lambda\Big(0.0945\Big(
	\frac{\tilde{\beta}^4}{\lambda^3}\Big)^{1/3}\!+\!0.0065
	\Big)\!-\!0.0253\frac{\lambda^{5/2}}{\tilde{\beta}^2}\Bigg(
	\Big(
		\frac{\tilde{\beta}^4}{\lambda^3}\Big)^{5/6}\!+\!0.9603
			\sqrt{\frac{\tilde{\beta}^4}{\lambda^3}}\Bigg)}{\tilde{
				\beta}^2
			\Big(3.7321\Big(\frac{\tilde{\beta}^4}{\lambda^3}\Big)^{1/3}
			\!+\!0.0515\Big)\!+\!\lambda^{3/2}
			\Big(\Big(
			\frac{\tilde{\beta}^4}{\lambda^3}\Big)^{5/6}\!+\!0.1921
			\sqrt{\frac{\tilde{\beta}^4}{\lambda^3}}\Big)}.
\end{eqnarray}
The non-perturbative $\langle\varphi\rangle$ is
\begin{eqnarray}
\langle\varphi\rangle\!=\!N^{\!-\!1}\int_{-\infty}^{\infty}
d\varphi \varphi\sum_{n\!=\!0}^{\infty}\frac{1}{n!}\Big(\!-\!\frac{4\pi^2\beta\varphi^3}{9H^4}\Big)^n
e^{\frac{\!-\!8\pi^2}{3H^4}(\frac{1}{2}m^2\varphi^2\!+\!\frac{\lambda}{4!}\varphi^4)},
\end{eqnarray}
where
\begin{eqnarray}
N\!=\!\int_{-\infty}^{\infty}d\varphi \sum_{n\!=\!0}^{\infty}\frac{1}{n!}\Big(\!-\!\frac{4\pi^2\beta\varphi^3}{9H^4}\Big)^n
 e^{\frac{\!-\!8\pi^2}{3H^4}(\frac{1}{2}m^2\varphi^2\!+\!\frac{\lambda}{4!}\varphi^4)}\;.
 \end{eqnarray}
Evaluating the integral, we find
 \begin{eqnarray}
 &&\langle\varphi\rangle\!=\!-N^{\!-\!1}\sum_{n\!=\!0}^{\infty}
 \frac{2^{4n\!+\!1}\pi^{n\!-\!\frac{1}{2}}\beta^{2n\!+\!1}}{(2n\!+\!1)!\lambda^{\frac{7\!+\!6n}{4}}H^{2n\!+\!1}3^{n\!-\!\frac{1}{2}}}
\Bigg(3m^2\pi\left(2n\!+\!1\right)\Gamma\Big[\frac{3(2n\!+\!1)}{4}\Big]{}_{1}\mathcal{F}_{1}\Big[\frac{6n\!+\!7}{4},
 \frac{3}{2},\frac{4m^4\pi^2}{H^4\lambda}\Big]\label{gulay2}
 \nonumber\\
 &&\hspace{3.5cm}\!-\!\sqrt{\lambda}H^2
 \Gamma\Big[\frac{6n\!+\!5}{4}\Big]{}_{1}\mathcal{F}_{1}\Big[\frac{6n\!+\!5}{4},
 \frac{1}{2},\frac{4m^4\pi^2}{H^4\lambda}\Big]\bigg),\label{gülay7}
 \end{eqnarray}
 where
 \begin{eqnarray}
 &&N\!=\!\sum_{n\!=\!0}^{\infty}\frac{2^{4n\!-\!1}\pi^{n\!-\!\frac{1}{2}}\beta^{2n}}{(2n)!
 	\lambda^{\frac{5\!+\!6n}{4}}H^{2n\!-\!1}3^{n\!-\!\frac{1}{2}}}\Bigg(\lambda
 \Gamma\Big[\frac{1\!+\!6n}{4}\Big]
 {}_{1}\mathcal{F}_{1}\Big[\frac{1\!+\!6n}{4},\frac{1}{2},\frac{4m^4\pi^2}{H^4\lambda}\Big]
 \!-\!4m^2\pi\frac{\sqrt{\lambda}}{H^2}
 \nonumber\\
 &&\hspace{4cm}\times\Gamma\Big[\frac{3(1\!+\!2n)}{4}\Big]
 {}_{1}\mathcal{F}_{1}\Big[\frac{3(1\!+\!2n)}{4},\frac{3}{2},
 \frac{4m^4\pi^2}{H^4\lambda}\Big]
 \bigg).
 \end{eqnarray}
 \begin{figure}
 	\centering
 	\includegraphics[width=11.5cm,height=6.5cm]{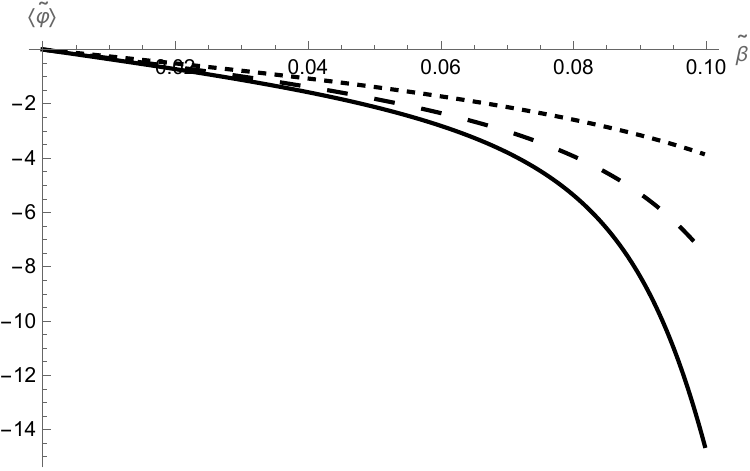}
 	\vspace*{2.256mm}
 	\caption{Plots of $\langle\tilde{\varphi}\rangle$ for the massive scalar, defined in Eq.~(\ref{gülay7}) versus $\tilde{\beta}$ for  different values of $\lambda$. The dashed, large-dashed and solid lines are for $\lambda\!=\!0.015, 0.01$ and $0.008$, respectively.}
 	\label{kemal6}
 \end{figure}

 In Fig$.\!\quad\!3$, as the coupling constant $\lambda$ increases, the vacuum expectation value of the massive scalar, 
$\langle \tilde{\varphi} \rangle$ increases.  
Conversely, as the cubic coupling constant $\tilde{\beta}$ increases, the vacuum expectation value of the massive scalar decreases.

 In the massless limit, we \cite{Joshi} obtain 
 \begin{eqnarray}
 &&\lim_{m\rightarrow0}\langle\varphi\rangle\!=\!-\frac{\beta}{\lambda}\Bigg({}_{2}\mathcal{F}_{2}\Big[\Big\{\frac{5}{12},\frac{13}{12}\Big\},\Big\{\frac{1}{2},\frac{5}{4}\Big\},
 \frac{3\pi^2\beta^4}{H^4\lambda^3}\Big]\!+\!\frac{14}{3}\pi\frac{\beta^2}{\lambda^2}
 \frac{\sqrt{\lambda}}{H^2}\frac{\Gamma\Big[\frac{3}{4}\Big]}{\Gamma\Big[\frac{1}{4}\Big]}
 \nonumber\\
 &&\times{}_{2}\mathcal{F}_{2}\Big[\Big\{\frac{11}{12},\frac{19}{12}\Big\},\Big\{\frac{3}{2},
 \frac{7}{4}\Big\},
 \frac{3\pi^2\beta^4}{H^4\lambda^3}\Big]
 \bigg)/\Bigg({}_{2}\mathcal{F}_{2}\Big[\Big\{\frac{1}{12},\frac{5}{12}\Big\},\Big\{\frac{1}{4},\frac{1}{2}\Big\},
 \frac{3\pi^2\beta^4}{H^4\lambda^3}\Big]
 \nonumber\\
 &&\hspace{2cm}\!+\frac{2\pi\beta^2\sqrt{\lambda}}{\lambda^2H^2}
 \frac{\Gamma\Big[\frac{3}{4}\Big]}{\Gamma\Big[\frac{1}{4}\Big]}
 {}_{2}\mathcal{F}_{2}\Big[\Big\{\frac{7}{12},\frac{11}{12}\Big\},\Big\{\frac{3}{4},
 \frac{3}{2}\Big\},
 \frac{3\pi^2\beta^4}{H^4\lambda^3}\Big]
 \bigg).\label{gülay8}
 \end{eqnarray}
  \begin{figure}
 	\centering
 	\includegraphics[width=11.5cm,height=6.5cm]{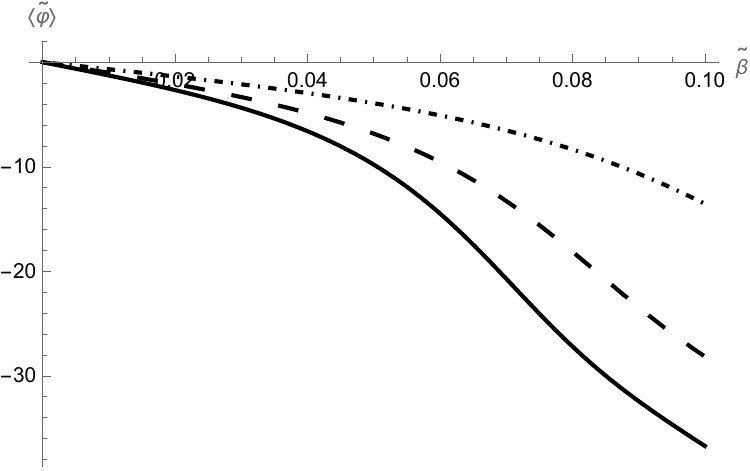}
 	\vspace*{2.256mm}
 	\caption{Plots of $\langle\varphi\rangle$ for the massless scalar, defined in Eq.~(\ref{gülay8}) versus $\bar{\beta}$ for  different values of $\lambda$. The dashed, large-dashed and solid lines are for $\lambda\!=\!0.015, 0.01$ and $0.008$, respectively.}
 	\label{kemal6}
 \end{figure}
 For small cubic coupling, the $\langle\tilde{\varphi}\rangle$ is
  \begin{eqnarray}
 	\langle\tilde{\varphi}\rangle\!=\!-\frac{\tilde{\beta}}{\lambda}\!-\!
 	2.8315\frac{\tilde{\beta}^3}{\lambda^{\frac{5}{2}}}\!+\!
 \mathcal{O}\left(\tilde{\beta}^4\right)\;,
 \end{eqnarray}
 where $\tilde{\beta}\!=\!\frac{\beta}{H}$, $\tilde{\varphi}\!=\!\frac{\varphi}{H}$
 and for large $\tilde{\beta}^4/\lambda^3\gg1$, Eq.~$\left(\ref{gülay8}\right)$ is obtained as
 \begin{eqnarray}
 \langle\tilde{\varphi}\rangle\!=\!-\frac{0.3915\times
 	\tilde{\beta}^{1/3}\Big(\tilde{\beta}^2
 	\left(\frac{\tilde{\beta}^4}{\lambda^3}\right)^{1/6}
 	\!+\!0.0239\lambda^{3/2}\Big)}{\lambda^{1/2}\Big[\tilde{\beta}^2\Big(0.268
 		\left(\frac{\tilde{\beta}^4}{\lambda^3}\right)^{1/3}\!+\!
 		0.0515\Big)\!+\!\lambda^{3/2}\Big(\left(
 		\frac{\tilde{\beta}^4}{\lambda^3}\right)^{5/6}\!+\!
 		0.0138\sqrt{\frac{\tilde{\beta}^4}{\lambda^3}}\Big)
 			\Big]}\;.
 \end{eqnarray}
$\langle\tilde{\varphi}\rangle$  is negative for all $ \tilde{\beta}>0$ and is independent of $H$.  When $\lambda\rightarrow0$, $\langle\tilde{\varphi}\rangle$ decreases unboundly so it diverges. Hence this must lead to instability.

For $V(\varphi)\!=\!(1/2)m^2\varphi^2$, we get the $\langle{\varphi}^2\rangle$,
\begin{eqnarray}
\langle{\varphi}^2\rangle\!=\!\frac{3H^4}{8\pi^2m^2}\;,
\end{eqnarray}
and the $\langle{\varphi}\rangle$ doesn't contribute at tree order.

\section{Quantum corrected correlators via quantum field theory}
\label{sec:stochastic}
The two-point correlation function of the infrared truncated massive minimally coupled spectator scalar field was computed at tree, one-and two-loop order applying stochastic formalism  with $\frac{1}{2}m^2\varphi^2\!+\!\frac{\lambda\varphi^4}{4!}$ potential in $\cite{GV}$. In this section, we compute the quantum corrected two-point correlation function with asymmetric potential. Firstly, varying the Lagrangian with metric~(\ref{desitter}), density (\ref{lagrange}) yields the scalar field equation
\begin{eqnarray}
\hspace{0.4cm}\ddot{\varphi}(t,\vec{x})\!+\!(D\!-\!1) H \dot{\varphi}(t,\vec{x})\!-\!\!
\left[\frac{\nabla^2}{a^2}-\!m^2\right]\!\!\varphi(t,\vec{x})\!=\!-\frac{V'(\varphi)(t,\vec{x})}{1\!\!+\!\delta Z}\; ,\label{feq}
\end{eqnarray}
where an overdot denotes the derivative with respect to comoving time $t$ and prime denotes the derivative with respect to the argument, hence
\begin{eqnarray}
\hspace{-0.2cm}V'(\varphi)(t,\vec{x})\!=\!\delta m^2
\varphi(t,\vec{x})\!+\!\frac{1}{6}(\lambda\!+\!\delta\lambda) \varphi^3(t,\vec{x})\!+\!\frac{1}{2}(\beta\!+\!\delta\beta)\varphi^2(t,\vec{x})\; .\label{potential}
\end{eqnarray}
Solution of Eq.~(\ref{feq}) can be given as
\begin{eqnarray}
\hspace{-0.4cm}\varphi(t,\vec{x})\!=\!\varphi_0(t,\vec{x})
\!-\!\!\int_0^t\!dt'a^{3}(t')\!\int\!d^{3}x'G(t,
\vec{x};t'\!, \vec{x}\,')\frac{V'(\varphi)(t',\vec{x}\,')}{1\!\!+\!\delta Z}\; ,
\label{fullfield}
\end{eqnarray}
where the free field $\varphi_0(t, \vec{x})$ in Eq.~(\ref{fullfield}) obeys
the homogeneous field equation and agrees with the interacting (full) field
at initial comoving time $t\!=\!t_I\!=\!0$. The free field can be expressed in terms of Hankel functions of the first and second kind. The Green's function ${G}(t,
\vec{x};t'\!, \vec{x}\,')$ in Eq.~(\ref{fullfield}), on the other hand, is any solution of the field equation with a Dirac-delta source term $\delta(t\!-\!t')\delta^{3}(\vec{x}\!-\!\vec{x}\,')$ which obeys the retarded boundary conditions. 
Spatially Fourier transformed free field equation is 
\begin{eqnarray}
{\widetilde{\ddot{\varphi}}}_0(t,\vec{k})\!+\!3 H {\widetilde{\dot{\varphi}}}_0(t,\vec{k})\!+\!\!
\left[\frac{k^2}{a^2}\!+\!m^2\right]\!{\widetilde{\varphi}}_0(t,\vec{k})\!=\!0 \; .\label{Fouriertransfieldeq}
\end{eqnarray}
To study the IR physics during inflation, one can cut out~\cite{Star,StarYok} the ultraviolet modes with wave number $k\!>\!Ha(t)$ in mode expansion by introducing a dynamical Heaviside step function $\Theta$ in Fourier space, 
\beeq
\varphi_0(t,\vec{x}) \!=\! H^{\frac{3}{2}}\! \sum_{\vec{n} \neq 0} \Theta(Ha(t)\!-\!k)\!\left[ u(t,k) e^{i \vec{k} \cdot \vec{x}}
\hat{A}_{\vec{n}} \!+\! u^*(t,k) e^{-i \vec{k} \cdot \vec{x}} \hat{A}^{\dagger}_{\vec{n}}
\right]\; .\label{truncexpwithu1}
\eneq
We obtain the leading IR limit of the mode function
\beeq
u(t, k)\longrightarrow\frac{1}{\sqrt{Ha^{3}(t)}}
\frac{\Gamma(2\nu)}{\Gamma(\nu\!+\!\frac{1}{2})}  \left(\!\frac{2k}{Ha(t)}\!\right)^{-\nu}\!\!\Biggl\{\!1\!\!+\!
\mathcal{O}\!\left(\!\Bigl(\frac{k}{H a(t)}\Bigr)^{2\beta}\right)\!\Biggr\}\; ,\label{modeleadingorder1}
\eneq
where $\nu\!=\!\sqrt{\frac{9}{4}\!-\!\frac{m^2}{H^2}}$ and $\beta\!=\!\nu\!$ if $\frac{1}{2}\sqrt{5}\!<\!\frac{m}{H}\!<\!\frac{3}{2}$ or $\beta\!=\!1$ if $\frac{m}{H}\!<\!\!\frac{1}{2}\sqrt{5}\!<\!\frac{3}{2}$.  Then, inserting $u(t, k)$ in Eq.~(\ref{modeleadingorder1}) into Eq.~(\ref{truncexpwithu1}) we find the IR truncated massive free field in $4$-dimensions,
\be
{\bar\varphi}_0(t, \vec{x})\!=\!H^{\nu+1}\!\frac{\Gamma(2\nu)}{\Gamma(\nu\!+\!\frac{1}{2})2^\nu}\,
a^{\nu-\frac{3}{2}}(t)\!\sum_{\vec{n} \neq 0} \!\frac{\Theta\left(H a(t) \!-\! k\right)}{k^{\nu}}\!\left[e^{i \vec{k} \cdot \vec{x}}\hat{A}_{\vec{n}} \!+\! e^{-i \vec{k} \cdot \vec{x}} \hat{A}^{\dagger}_{\vec{n}}
\right] \; .\label{massivefree}
\ee
The commutator function $\left[\varphi_0(t,\vec{x}), \varphi_0(t'\!,\vec{x}\,')\right]$ provides a convenient representation for the Green's function,\beeq
G(t,\vec{x};t'\!, \vec{x}\,')\!=\!i\Theta(t\!-\!t')\left[\varphi_0(t,\vec{x}), \varphi_0(t'\!,\vec{x}\,')\right]\; .\label{commutatorgreen}
\eneq
 To get the IR truncated full field 
 $\bar{\varphi}(t,\vec{x})$, using Eqns. $(\ref{massivefree}-\ref{commutatorgreen})$ the IR limit of Green's function~(\ref{commutatorgreen}) at leading order is obtained as 
\begin{eqnarray}
G(t,\vec{x};t'\!, \vec{x}\,')\!\longrightarrow\!\frac{\Theta(t\!-\!t')}{2H\nu}\Biggl[\frac{a^{2\nu}(t)\!-\!a^{2\nu}(t')}
{[a(t)\,a(t')]^{\nu+\frac{3}{2}}}\Biggr]\delta^{3}(\vec{x}\!-\!\vec{x}\,')\; .\label{GreenIRlimit}
\end{eqnarray}
Substituting limit~(\ref{GreenIRlimit}) into Eq.~(\ref{fullfield}) yields the IR truncated full field as
\be
\bar\varphi(t,\vec{x}) \!=\! \bar\varphi_0(t,\vec{x})
\!-\!\frac{1}{2\nu H}\!\int_0^t \!\!dt' \!\left[\left(\frac{a(t)}{a(t')}\right)^{\nu-\frac{3}{2}}
\!\!-\!\left(\frac{a(t')}{a(t)}\right)^{\nu+\frac{3}{2}}\right]\!\frac{V'(\bar\varphi)(t'\!,\vec{x})}{1\!\!+\!\delta Z} \; ,\label{fullfieldgeneral}
\ee
where the potential is as given Eq.~(\ref{potential}). The latter term in the square brackets can be neglected next to the former which dominates throughout the range of integration. Moreover, the counterterms in potential~(\ref{potential}) cannot contribute \cite{Wstocsqed} in the leading order we consider. One can see this by comparing the powers of the fields and orders of $\lambda$ involved in various counterterms in the model: $\delta \lambda \!\sim\! \mathcal{O}(\lambda^2)$, $\delta m^2 \!\sim\! \mathcal{O}(\lambda)$ and $\delta Z \!\sim\! \mathcal{O}(\lambda^2)$ \cite{BOW}. Firstly, compare the contributions involving $\lambda\varphi^4$ and the contributions involving $\delta\lambda\varphi^4$ terms. Powers of the fields are the same, so the former and the latter have the same structure of leading terms. The latter, however, are suppressed by at least one extra factor of $\lambda$ (with $\delta \lambda \!\sim\! \mathcal{O}(\lambda^2)$), they can never be in leading order. Secondly, compare the contributions involving $\lambda\varphi^4$ and the contributions involving $\delta m^2\varphi^2$ terms. Although the former and the latter are linear in $\lambda$ (with $\delta m^2 \!\sim\! \mathcal{O}(\lambda)$), the former are quartic in field whereas the latter are quadratic in field. Therefore, at a given order in $\lambda$, the latter can never have as high order leading terms as the former. Finally, the field strength counterterm $\delta Z$ appears in Eq.~(\ref{fullfieldgeneral}) in the form
\be
\frac{V'(\bar\varphi)(t'\!,\vec{x})}{1\!+\!\delta Z}\!=\!V'(\bar\varphi)\left[1\!-\!\delta Z\!+\!(\delta Z)^2\!-\!\cdots\right]\; ,
\ee
with $\delta Z \!\sim\! \mathcal{O}(\lambda^2)$. Hence, exactly the same leading order contributions that Eq.~(\ref{fullfieldgeneral}) would yield are obtained from its simplified version without the counterterms, i.e., from
\begin{eqnarray}
\bar\varphi(t,\vec{x}) \!=\! \bar\varphi_0(t,\vec{x})
\!-\!\frac{a^{-\frac{\delta}{2}}(t)}{(2\nu)H}\,\!\!\int_0^t \!\!dt'a^{\frac{\delta}{2}}(t')\,\Big[\frac{1}{6}\lambda{\bar\varphi}_0^3(t',\vec{x})\!+\!\frac{1}{2}
\beta{\bar\varphi}_0^2(t',\vec{x})\Big]
 \; ,\label{freefieldsimple}
\end{eqnarray}
where we define \begin{eqnarray}
3\!-\!\!2\nu\!\equiv\delta\; .\label{delta}
\end{eqnarray}
(Note that $\delta\!\!\rightarrow\!0$ as the mass $m\!\!\rightarrow\!0$.) Infrared truncated full field $\bar\varphi(t,\vec{x})$ can be expressed in terms of the IR truncated free field $\bar\varphi_0(t,\vec{x})$, at any order of $\lambda$ and $\beta$, by iterating Eq.~(\ref{freefieldsimple}) successively. Iterating it twice, the two-point correlation function of the IR truncated full field for two distinct events
\begin{eqnarray}
&&\hspace{1.2cm}\langle\Omega|
\bar{\varphi}(t,\vec{x})\bar{\varphi}(t'\!,\vec{x}\,')|\Omega\rangle\!\!=\!\!\langle\Omega|
\bar{\varphi}_0(t,\vec{x})\bar{\varphi}_0(t'\!,\vec{x}\,')|\Omega\rangle\!-\!\frac{a^{\nu\!-\!\frac{3}{2}}}{2\nu H}\int_{0}^{t}dt''a''^{\frac{3}{2}\!-\!\nu}\nonumber\\
&&\hspace{0.1cm}\times\Bigg[\frac{\lambda}{3!}\langle\Omega|
{\bar{\varphi}}_{0}^3(t'',\vec{x})\bar{\varphi}_{0}(t'\!,\vec{x}\,')|\Omega\rangle\!+\!\frac{\beta}{2!}\langle\Omega|
{\bar{\varphi}}_{0}^2(t'',\vec{x})\bar{\varphi}_{0}(t'\!,\vec{x}\,')|\Omega\rangle\Bigg]\!-\!\frac{a'^{\nu\!-\!
		\frac{3}{2}}}{2\nu H}\int_{0}^{t'} dt''a''^{\frac{3}{2}\!-\!\nu}\nonumber
\\
&&\hspace{0.6cm}\times\Bigg[\frac{\lambda}{3!}\langle\Omega|
{\bar{\varphi}}_{0}^3(t'',\vec{x}\,')\bar{\varphi}_{0}(t'\!,\vec{x})|\Omega\rangle\!+\!
\frac{\beta}{2!}\langle\Omega|
{\bar{\varphi}}_{0}^2(t'',\vec{x}\,')\bar{\varphi}_{0}(t,\vec{x})|\Omega\rangle\Bigg]\!+\!
\mathcal{O}(\lambda^2,\beta^2,\lambda\beta)\;.
\end{eqnarray}
At order $\beta$, the one-loop contributions to the two-point correlation function for the massive scalar field vanish.  Here, they contribute only at order $\lambda$.

The vacuum expectation value of the IR truncated full field is given as
\begin{eqnarray}
\langle\bar\varphi(t,\vec{x})\rangle \!=
-\frac{a^{-\frac{\delta}{2}}(t)}{(2\nu)H}\,\!\!\int_0^t \!\!dt'a^{\frac{\delta}{2}}(t')\,\!\frac{\beta}{2}
\langle{\bar\varphi}_0^2(t',\vec{x})\rangle\!+\!\mathcal{O}\left(\lambda\beta\right)
 \; .\label{freefieldsimple2}
\end{eqnarray}
It follows from the above equation that $\langle\bar{\varphi}\rangle$ does not contribute at $\mathcal{O}\left(\lambda\right)$.
\section{Two-point correlation function}
\label{sec:twopointcorrelator}

The two-point correlation function of the IR truncated full field for two distinct events
\begin{eqnarray}
\hspace{-0.4cm}\langle\Omega|
\bar{\varphi}(t,\vec{x})\bar{\varphi}(t'\!,\vec{x}\,')|\Omega\rangle\!=\!\langle\Omega|
\bar{\varphi}(t,\vec{x})\bar{\varphi}(t'\!,\vec{x}\,')|\Omega\rangle_{\rm tree}\!&+&\!\langle\Omega|
\bar{\varphi}(t,\vec{x})\bar{\varphi}(t'\!,\vec{x}\,')|\Omega\rangle_{\rm 1-loop}\nonumber\\
&+&\!\mathcal{O}(\lambda^2,\beta^2,\lambda\beta)\;,\label{genelfullexpect}
\end{eqnarray}
with $t'\!\leq\!t$ and $\vec{x}\,'\!\!\neq\!\vec{x}$, can be obtained for the field with an asymmetric self-interaction. It yields, at tree-order,
\begin{eqnarray}
\langle\Omega|
\bar{\varphi}(t,\vec{x})\bar{\varphi}(t'\!,\vec{x}\,')|\Omega\rangle_{\rm tree}\!=\!\langle\Omega|
\bar{\varphi}_0(t,\vec{x})\bar{\varphi}_0(t'\!,\vec{x}\,')|\Omega\rangle\; .
\end{eqnarray}
The leading (one-loop) quantum correction
\begin{eqnarray}
&&\hspace{0cm}\langle\Omega|                                                                
\bar{\varphi}(t,\vec{x})\bar{\varphi}(t'\!,\vec{x}\,')|\Omega\rangle_{\rm 1-loop}\!=\!-\frac{\lambda}{6(2\nu)H}\Bigg[a^{-\frac{\delta}{2}}(t')\langle\Omega|
\bar{\varphi}_0(t,\vec{x})\!\!\!\int_0^{t'}\!\!\!\!d\tilde{t}\,a^{\frac{\delta}{2}}(\tilde{t})
\,{\bar{\varphi}}^3_0(\tilde{t},\vec{x}\,')
|\Omega\rangle\nonumber\\
&&\hspace{3cm}+a^{-\frac{\delta}{2}}(t)\langle\Omega|\!\!
\int_0^{t}\!\!\!dt''a^{\frac{\delta}{2}}(t''){\bar{\varphi}}^3_0(t''\!,\vec{x})
\bar{\varphi}_0(t'\!,\vec{x}\,')|\Omega\rangle\Bigg]\; ,\label{1loopcorr}
\end{eqnarray}
is not hard to compute,
where $0\!\leq\!t''\!\leq\!t$ and $t'\!\leq\!t$, however, is demanding because it involves three VEVs each of which has a double time integral without a definite time ordering in the integrand.  Note also that perturbation theory breaks down when $\ln(a(t))\!=\!Ht\!\sim\!1/\sqrt{\lambda}$ \cite{OW1}. 

The tree-order two-point correlation function of the IR truncated massive scalar field\cite{GV} is obtained
using Eqs.~(\ref{massivefree}) and (\ref{delta}) as
\begin{eqnarray}
\langle\Omega|{\bar\varphi}_0(t,\vec{x}) {\bar\varphi}_0(t'\!,\vec{x}\,')|\Omega\rangle\!\simeq\!
\mathcal{A}\,\emph{f}_0(t,t'\!,\Delta x)\; ,\label{treecorrgamma}
\end{eqnarray}
where the constant
\begin{eqnarray}
\mathcal{A}\!\equiv\!\frac{\Gamma^2\!\left(2\nu\right)}{\Gamma^2\!\left(\nu\!+\!\frac{1}{2}\right)}
\frac{H^{2}}{2^{2\nu}\pi^{2}}\;,\label{Acons}
\end{eqnarray}
and the spacetime and mass dependent function\begin{eqnarray}
&&\hspace{-0.5cm}\emph{f}_0(t,t'\!,\Delta x)\!=\!\!\frac{1}{2}\!\left[-\alpha(t)\,\alpha(t')\right]^{-\frac{\delta}{2}}
\!\Bigg\{\!\Gamma\!\left(-\!1\!\!+\!\delta, i\alpha(t')\right)\!-\!\Gamma\!\left(-\!1\!\!+\!\delta, iH\!\Delta x\right)\!+\!(-1)^{-\delta}\Big[\Gamma\!\left(-\!1\!\!+\!\delta, -i\alpha(t')\right)\nonumber\\
&&\hspace{5.6cm}-\Gamma\!\left(-\!1\!\!+\!\delta, -iH\!\Delta x\right)\!\Big]\!\Bigg\}\; ,\label{treecorrgammaf}
\end{eqnarray}
with
\beeq
\alpha(t)\!\equiv\!\alpha(t,\Delta x)\!\equiv\!a(t)H\!\Delta x\; .\label{alphaoft}
\eneq
where $t'\!\!\leq\!t$, $\vec{x}\,'\!\neq\!\vec{x}$ and $\Delta x\equiv\parallel\!\Delta\vec{x}\!\parallel=
\parallel\!\vec{x}\!-\!\vec{x}\,'\!\parallel$.

Employing power series representation of the incomplete gamma function in Eq.~(\ref{treecorrgammaf}) we obtain
\begin{eqnarray}
\emph{f}_0(t,t'\!,\Delta x)\!=\!\Big[a(t)\,a(t')\Big]^{-\frac{\delta}{2}}\!\sum_{n=0}^\infty\!\frac{(-1)^n (H\!\Delta x)^{2n}}{(2n\!\!+\!\!1\!)!} \frac{a^{2n+\delta}(t')\!-\!\!1}{2n\!\!+\!\delta}\; .
\label{treecorrseries}
\end{eqnarray}
Taking the equal space-time limit of Eq.~(\ref{treecorrseries}) leads to \beeq
\langle\Omega|\bar{\varphi}^2_0(t,\vec{x})|\Omega\rangle\!\simeq\!
\mathcal{A}
\frac{1\!\!-\!a^{-\delta}(t)}
{\delta}\; .\label{phisquare}
\eneq 

Eqs.~(\ref{treecorrgamma})-(\ref{treecorrseries}) yield the tree-order correlator for an IR-truncated {\it massive} scalar field in both analytic function and power series forms.  
We use this tree-order correlator in the perturbative computation of the quantum-corrected correlation function in a self-interacting theory.

Computation of one-loop contribution~(\ref{1loopcorr}) to two-point correlation~(\ref{genelfullexpect}) involves evaluations of two VEVs. 
The one-loop correlator for the massive scalar is obtained as
\beeq
\langle\Omega|
\bar{\varphi}(t,\vec{x})\bar{\varphi}(t'\!,\vec{x}\,')|\Omega\rangle_{\rm 1-loop}\!\simeq\!-\frac
{\lambda}{2\nu}\frac{{\mathcal{A}}^2}{H^2}\,\emph{f}_{1}(t,t'\!,\Delta x)\;,\label{1loopmassivefinal}
\eneq
where we define the spacetime and mass dependent function\be
&&\hspace{-0.45cm}\emph{f}_1(t,t'\!,\Delta x)\!=\!\!\frac{\left[a(t)\,a(t')\right]^{-\frac{\delta}{2}}}
{2\,\delta}\!\!\sum_{n=0}^\infty
\!\frac{(-1)^{n}(H\!\Delta x)^{2n}}{(2n\!\!+\!\!1\!)!(2n\!\!+\!\delta)}
\Bigg\{\!a^{2n}(t')\!
\Bigg[a^{\delta}(t')\!\Bigg(\!\!\ln(a(t))\!-\!\ln(a(t'))\!\!+\!\frac{a^{-\delta}(t)}{\delta}\!\Bigg)
\!\!-\!\frac{1}{\delta}\Bigg]\nonumber\\
&&\hspace{0.65cm}-\ln(a(t))\!-\!\ln(a(t'))
\!+\!\frac{1\!\!-\!a^{-\delta}(t)}{\delta}\!+\!\frac{1\!\!-\!a^{-\delta}(t')}{\delta}
\!-\!2\!\left[\frac{1\!-\!a^{2n+\delta}(t')}{2n\!\!+\!\delta}\right]\!\!+\!
\frac{1\!\!-\!a^{2n}(t')}{n}\!\Bigg\}\; .\label{fmassive}
\ee
Using the equal spacetime limit of Eq.~(\ref{fmassive}) in Eq.~(\ref{1loopmassivefinal}) yields the VEV of the field strength squared
\beeq
\langle\Omega|
\bar{\varphi}^2(t,\vec{x})|\Omega\rangle_{\rm 1-loop}\!\simeq\!-\frac
{\lambda}{2\nu}\frac{{\mathcal{A}}^2}{H^2}
\frac{a^{-\delta}(t)}{\delta^2}
\left[\frac{a^{\delta}(t)\!-\!a^{-\delta}(t)}{\delta}\!-\!2\ln(a(t))\right]\;.\label{1loopeqlspctm}
\eneq
A constraint on the coupling constant $\lambda$ can immediately be deduced here. Magnitude of one-loop correction~(\ref{1loopeqlspctm}) ought to remain less than the magnitude of tree-order correlator~(\ref{phisquare}) for the perturbation theory to be valid. This implies that the inequality\begin{eqnarray}
\lambda\!<\!\frac{H^2}{\mathcal{A}}2\nu\delta\!\left[\frac{1\!+\!a^{-\delta}(t)}{\delta}
\!-\!\frac{2\ln(a(t))}{a^{\delta}(t)\!-\!1}\right]^{-1}\; ,\label{lambda}
\end{eqnarray}
must hold during inflation. Massless limit of Eq.~(\ref{lambda}) yields $\lambda\!<\!36\pi^2/\ln^2(a(t))$ in $D\!=\!4$,   in agreement with the note---stated in Sec.~\ref{sec:twopointcorrelator}---that the perturbation theory breaks down when $\ln(a(t))\!\sim\!1/\sqrt{\lambda}$. Note also that the massless limit of Eq.~(\ref{fmassive}) yields the one-loop correlator for the massless scalar.

Using Eq. $\left(\ref{freefieldsimple2}\right)$, we obtain the vacuum expectation value of the infrared truncated massive minimally scalar field as
\begin{eqnarray}
\langle\bar\varphi(t,\vec{x})\rangle \!=
-\frac{\beta}{2^{2\nu\!+\!1}\nu\pi^2}\frac{\Gamma^2\left(2\nu\right)}{\Gamma^2\left(\nu\!+\!\frac{1}{2}\right)}
\frac{\left(1\!-\!a^\frac{\!-\!\delta}{2}(t)\right)^2}{\delta^2}\!+\!\mathcal{O}\left(\lambda\beta\right)\;.
\end{eqnarray}
This shows that there is no contribution at tree level; contributions appear only at one-loop order at $\mathcal{O}(\beta)$.  At late times, we obtain $\langle\bar{\varphi}\rangle$ 

\begin{eqnarray}
\langle\bar\varphi(t,\vec{x})\rangle \!=
-\frac{\beta}{2^{2\nu\!+\!1}\nu\pi^2}
\frac{\Gamma^2\left(2\nu\right)}{\Gamma^2\left(\nu\!+\!\frac{1}{2}\right)}
\frac{1}{\delta^2}\!+\!\mathcal{O}\left(\lambda\beta\right)\;.
\end{eqnarray}
In the massless case $\left(m=0, \, \delta=0\right)$, the $\langle\bar\varphi(t,\vec{x})\rangle$ is
\begin{eqnarray}
\langle\bar\varphi(t,\vec{x})\rangle\!=-\frac{\beta}{2^3 3\pi^2}\ln^2(a(t))\!+\!\mathcal{O}\left(\lambda\beta\right)\;.
\end{eqnarray}
We see that as the scale factor $a(t)$ increases, $\langle\bar\varphi(t,\vec{x})\rangle$ decreases. 
 \section{DISCUSSION AND CONCLUSIONS}  
\label{sec:conclusions}
  We considered an infrared-truncated, massive, minimally coupled scalar field during inflation. 
In this work, we computed the equal-spacetime two-point correlation function using two different methods: the Fokker--Planck equation and perturbative quantum field theory.

First, we used the Fokker--Planck equation to obtain the non-perturbative vacuum expectation value of $\varphi$ and the coincident correlator $\langle \varphi^{2} \rangle$. 
When we calculate the vacuum expectation value of $\varphi$, it takes negative values, which may lead to a dynamical reduction of the inflationary cosmological constant at late times. 
Since $V(\varphi)$ is bounded from below for any value of $\beta$, we expect an equilibrium state at late times. 
As the coupling constant $\lambda$ increases, $\langle \varphi \rangle$ grows, but as the cubic coupling constant $\beta$ increases, $\langle \varphi \rangle$ decreases. For the one-loop--corrected two-point correlation function, we find that as $\lambda$ increases, $\langle \varphi^{2} \rangle$ decreases, while an increase in the cubic coupling constant $\beta$ causes the coincident correlator to increase.

Second, we evaluated the quantum-corrected two-point correlation function using stochastic formalism applied to perturbative quantum field theory. 
Here, as the coupling constant increases, the coincident correlator decreases. 
As the mass increases, the suppression becomes stronger; in fact, the one-loop correlator asymptotically approaches zero for masses larger than $H/2$.

Although the coincident correlators obtained by the two methods exhibit similar qualitative behavior, they differ numerically. 
We observe that as the cubic coupling constant increases, the difference between the two results grows. 
While quantum field theory yields a contribution to $\langle \varphi \rangle$ only at order $\beta$ at one-loop order, the Fokker--Planck approach predicts a contribution at order $\beta / \lambda$.

Despite these quantitative differences, the equal-spacetime two-point correlation functions obtained by both methods show consistent qualitative behavior. 
Both approaches reveal identical decay patterns and asymptotic scaling at large separations, reflecting the same underlying physical dynamics. 
This qualitative agreement---despite differing computational frameworks and assumptions---confirms the robustness of these methods in capturing the essential features of correlation functions during inflation. 
Consequently, the observed numerical discrepancies do not undermine the physical reliability of either approach but rather highlight the complementary insights they provide into the stochastic dynamics of the inflationary field.

Overall, the complementary nature of these two methods enriches our toolkit for studying inflationary physics. 
The intuitive and computationally accessible stochastic approach can guide analytical understanding and model building, while the probabilistic rigor of the Fokker--Planck equation offers precision and a pathway to exact solutions in tractable cases.

 \begin{acknowledgments}
I thank Vakıf Kemal Önemli and Nazmi Postacıoğlu for helpful comments.
\end{acknowledgments}

\end{document}